# Tunable Tesla-scale magnetic attosecond pulses through ring-current gating

Alba de las Heras,[1,*] Franco P. Bonafé,[2] Carlos Hernández-García,[1] Angel Rubio,[2,3,4,†] and Ofer Neufeld[2,‡]

[1]*Grupo de Investigación en Aplicaciones del Láser y Fotónica,*
*Departamento de Física Aplicada, Universidad de Salamanca, Salamanca 37008, Spain*
[2]*Max Planck Institute for the Structure and Dynamics of Matter and*
*Center for Free-Electron Laser Science, Hamburg 22761, Germany*
[3]*Center for Computational Quantum Physics, The Flatiron Institute, New York 10010, USA*
[4]*Nano-Bio Spectroscopy Group, Departamento de Física de Materiales,*
*Universidad del País Vasco, San Sebastián 20018, Spain*
(Dated: September 7, 2023)

Coherent control over electron dynamics in atoms and molecules using high-intensity circularly-polarized laser pulses gives rise to current loops, resulting in the emission of magnetic fields. We propose and demonstrate with ab-initio calculations "current-gating" schemes to generate direct or alternating-current magnetic pulses in the infrared spectral region, with highly tunable waveform and frequency, and showing femtosecond-to-attosecond pulse duration. In optimal conditions, the magnetic pulse can be highly isolated from the driving laser and exhibits a high flux density ($\sim 1$ Tesla at few hundred nanometers from the source, with a pulse duration of 787 attoseconds) for application in forefront experiments of ultrafast spectroscopy. Our work paves the way toward the generation of attosecond magnetic fields to probe ultrafast magnetization, chiral responses, and spin dynamics.

Ultrafast photoionization occurs in the interaction of strong laser pulses with atoms or molecules. In the low-frequency regime, the dominant strong-field ionization mechanism is the tunneling of bound electrons through the Coulomb potential barrier distorted by the laser field[1, 2]. The process of tunnel ionization has been extensively investigated in recent decades[3–8]. Besides its fundamental importance for understanding ultrafast laser-induced electron dynamics[9], it plays a crucial role in high-order harmonic generation (HHG) and attosecond science in atoms and molecules[10–12]. Tunnel ionization triggered by a circularly polarized driver is an especially intriguing process, since the tunneling barrier rotates during the laser interaction, inducing additional angular momenta on both the bound and continuum electron wavepackets[13–20]. As a result, the circularly-polarized laser pulse induces a long-lived current ring on the atomic scale[17, 21], leading to the generation of a long-lived magnetic field of high intensity[22, 23].

The conventional generation of strong magnetic fields typically relies on conducting or superconducting materials arranged into loops or coils where the electric current flows[24]. Such magnetic sources are particularly effective in quasi-static magnetic phenomena, or long-duration magnetization dynamics[25]. Also, more sophisticated strategies have been applied to improve the field strength, reaching up to 100 T in a magnetic texture[26]. However, these approaches do not provide a straightforward way to generate ultrashort magnetic pulses, which could initiate and probe ultrafast spin responses, chirality, or magnetic dynamics at high temporal resolution.

Alternatively, optical methods based on femtosecond laser pulses have made substantial progress in producing shorter magnetic bursts through laser-induced currents in plasmas[27–30], plasmonic nanostructures[31, 32], semiconductors[33–37], molecules[22, 23, 38–41] or atoms[42]. Interestingly, additional control over the magnetic field features can be gained by shaping the spatial structure of the laser beams[29, 30, 34–37, 43, 44]. Nevertheless, to our knowledge, the perspective of attosecond timescales has only been theoretically foreseen in relation to the attosecond electromagnetic pulses[22, 23, 39], where the dominance of the electric field overshadows the magnetic component. Therefore, the generation of isolated, strong, ultrashort magnetic fields is highly desirable to advance the control of purely magnetic phenomena, spintronics, chirotronics[45–51], or even HHG[52, 53]. Hence, the scenario of attosecond magnetic pulses would open up new avenues for the study and manipulation of magnetic phenomena on ultrafast timescales by accessing the fastest magnetic, spin, and chiral dynamics[44, 48, 54–58]. For instance, magnetic sources reaching the attosecond scale could boost scientific and technological breakthroughs in ultrafast magnetometry, magnetic phase transitions, materials science, plasma physics, topological systems, and high-speed data storage. However, many challenges in obtaining bright ultrashort tunable magnetic pulses remain unresolved, such as addressing macroscopic effects, separating the magnetic pulse from the electric field, maintaining a high flux far from the target, or controlling the timescales and wavelength.

In this letter, we propose laser-induced "current-gating" as a method to generate magnetic pulses within the femtosecond to attosecond regime. The long-lived electronic ring currents can be gated by using time-delayed counterrotating circularly-polarized laser pulses for high control over the pulse duration, waveform, and spectral content of the magnetic field emission. We demonstrate this proposal by performing ab-initio calculations of time-dependent density-functional theory (TDDFT) coupled to Maxwell equations[59, 60]. This state-of-the-art theoretical approach allows us not only to analyze the ultrafast non-perturbative and out-of-equilibrium multielectron dynamics occurring during the laser-atom interaction, but also the spatial and temporal profile of the magnetic field emission. We couple this ab-initio single-emitter methodology with an analytic model describing the response of a full macroscopic gas jet, as employed in experiments. Our theory predicts that under a suitable choice of the driving parameters, the generated magnetic field can be isolated from the driving laser, showing a high flux density of ∼1 Tesla at few hundred nanometers from a macroscopic target. We further demonstrate that by employing sets of more than two gating pulses, the waveform of the generated magnetic pulse can be controlled to mimic a few-cycle pulse with the desired wavelength and duration. Our results set a route toward the tunable generation of Tesla-scale attosecond magnetic field pulses.

---

* albadelasheras@usal.es
† angel.rubio@mpsd.mpg.de
‡ ofer.neufeld@gmail.com

We first outline the system of study and our methodology to describe the laser-atom interplay. Here, we explore the interaction of a train of intense circularly-polarized laser pulses with an atomic noble gas jet in a collinear setup. The laser carrier wavelengths (800 nm, 400 nm, 267 nm covering the infrared, visible and ultraviolet regions of the electromagnetic spectrum) are chosen to be off-resonant with the typical energy scales of the atom. Then, we are interested in high driving intensities to trigger highly nonlinear optical phenomena and strong-field ionization. We consider 4-cycle pulses full duration with a sinusoidal envelope, explicited in the supplemental material (SM). Theoretically, the multielectron dynamics during the light-matter interaction are described ab-initio within the time-dependent Kohn-Sham (KS) equations [61] of TDDFT in the length gauge

$$i\frac{\partial}{\partial t}|\varphi_n^{KS}(t)\rangle = \left[-\frac{1}{2}\nabla^2 + v_{KS}(\mathbf{r},t)\right]|\varphi_n^{KS}(t)\rangle \quad (1)$$

in the open-access OCTOPUS code[60]. The equations are given in atomic units, discretized over a real-space grid, and solved in real-time. $\varphi_n^{KS}(\mathbf{r},t)$ are the KS orbitals (with $n$ the orbital index), $c$ is the speed of light in vacuum, and $\mathbf{E}(t)$ is the external electric field in the dipole approximation (neglecting the magnetic field of the incoming laser pulse, which is a $c$ factor weaker than the electric field). The KS potential $v_{KS}(\mathbf{r},t) = v_{ion}(\mathbf{r}) + v_H(\mathbf{r},t) + v_{xc}[n(\mathbf{r},t)] - \mathbf{r}\cdot\mathbf{E}(t)$ includes the usual terms describing the electron-nuclei interaction $v_{ion}$ (which also incorporates interactions with deeper core electrons), the Hartree potential $v_H$, the exchange-correlation functional $v_{xc}$ of the electron density $n(\mathbf{r},t) = \sum_n |\langle\mathbf{r}|\varphi_n^{KS}(t)\rangle|^2$ [62], and the dipole term $-\mathbf{r}\cdot\mathbf{E}(t)$ describing the interaction with the laser field. We employ the adiabatic local density approximation including a self-interaction correction[63], and the frozen-core approximation for inner electrons by using norm-conserving pseudopotentials[64] for faster computational performance. Within this theoretical framework, the relevant observable for our study is the microscopic current density:

$$\mathbf{j}(\mathbf{r},t) = \frac{1}{2}\varphi_n^{KS,*}(\mathbf{r},t)\left(-i\boldsymbol{\nabla} + \frac{\mathbf{A}(t)}{c} - i[V_{ion},r]\right)\varphi_n^{KS}(\mathbf{r},t). \quad (2)$$

All additional details about the methodology and technical aspects can be found in the SM, where we also provide results from all-electron calculations or KS equations in the velocity gauge for further validation of our approach.

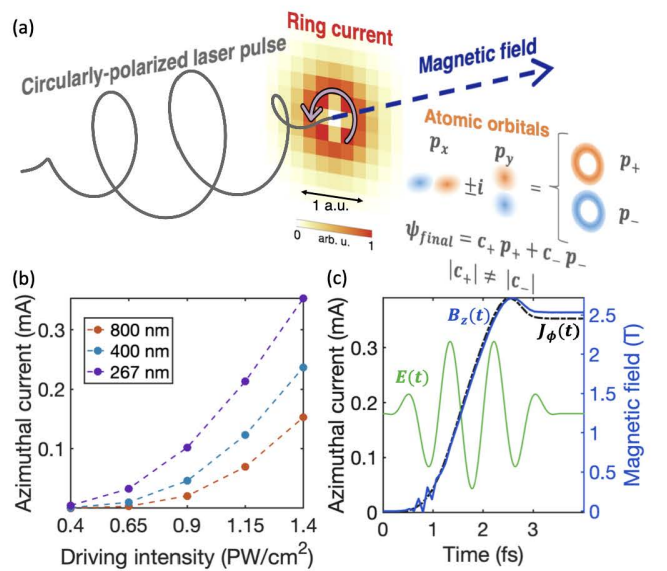

FIG. 1. (a) Scheme of a circularly-polarized electric pulse inducing a long-lived stationary azimuthal current, whose origin is associated with a different electron population in the $p_+$ and $p_-$ valence orbitals. The resulting magnetic field is static, linearly polarized along the longitudinal axis, and it persists after the laser pulse. (b) Dependence of the long-lived current amplitude on the driving peak intensity for different incident central wavelengths (267 nm is displayed in purple, 400 nm in blue, and 800 nm in orange). (c) The ultrafast build-up of the ring current (black dashed line) and the magnetic field (blue solid line) at the center of the current loop for a driving wavelength of 267 nm and peak intensity of 1.4 PW/cm². We also depict in (c) the temporal profile of the incoming circularly polarized electric pulse (green line in an arbitrary vertical axis).

Among the noble gases, we focus the study on neon atoms due to their electron configuration $1s^22s^22p^6$, which constitutes the most compact distribution of closed-shell valence p-orbitals. Figure 1(a) illustrates the scheme of a single intense off-resonant circularly-polarized electric laser pulse interacting with a neon atom. The figure shows the spatial profile of the current density obtained in our TDDFT calculations after the laser-atom interaction. The azimuthal current emerg-ing in the vicinity of the nuclei is the consequence of asymmetric electron populations in $p_+$ and $p_-$ valence orbitals, which arise from a propensity in ionization rates from these orbitals under a circularly-polarized driving laser[13–17]. Curiously, the direction of the current flow opposes the driving laser field because the ionization probability is higher for counter-rotating electrons than co-rotating electrons [13–18]. Overall, the resulting magnetic field is linearly polarized along the light propagation axis, and its orientation depends on the clockwise or anticlockwise direction of the current. In Fig.1(b), we present the dependence of the generated current amplitude on the driving peak intensity for different central wavelengths (267 nm, 400 nm, and 800nm). Our results demonstrate that the azimuthal current increases for drivers with higher peak intensity and shorter wavelength. This follows the expected trend, since the mechanism behind the ring current generation relies on strong-field ionization, and its amplitude is proportional to the total ionized charge.

After reproducing the appearance of stationary ring currents with our theoretical approach, we explore their build-up and their connection to magnetic field emission. Differently from previous approaches to obtain the magnetic field based on particle-in-cell codes[30, 35], the induced magnetic moment[43], or generalized Biot-Savart laws [36] like the Jefimenko equation[22, 39], we couple TDDFT and the microscopic Maxwell equations via the Riemann–Silberstein representation[59] using the multi-system framework of OCTOPUS[60]. This formalism allows a reformulation of Maxwell's equations in Schrödinger-like form, resulting in efficient coupled propagation in real-time and real-space of both the TDDFT KS equations and Maxwell's equations[59, 60].

In our numerical multi-system coupling, the current density computed via TDDFT is the input for the Maxwell solver. Our approach neglects the nondipole magnetic field effects and the atomic back-reaction, whose effect is overshadowed by the strong driving electric field. We show in Fig.1(c) the ultrafast turn-on preceding the stationary regime for a driving wavelength of 267 nm and peak intensity of 1.4 PW/cm$^2$. Noticeably, we certify a direct correspondence between the azimuthal stationary current (black dashed line) and the magnetic field (blue solid line), as expected from the well-established Ampère-Maxwell law: $\nabla \times \mathbf{B}(\mathbf{r},t) = \mu_0 \mathbf{j}(\mathbf{r},t) + \frac{1}{c^2}\frac{\partial \mathbf{E}(\mathbf{t})}{\partial t}$. Practically, it allows us to simplify the following analyses by only computing the electronic current, and generalizing the interpretations to the magnetic field. We refer to the SM for all numerical and technical details for this propagation scheme, as well as an in-depth analysis of the spatial distribution of the generated magnetic field. The temporal waveform of the driving circularly-polarized electric pulse is depicted in arbitrary units (green line in Fig.1 (c)) as a reference for readers.

Up until this stage, our results fully validate previous works on magnetic fields and impulses generated by strong-field-driven currents in other systems [30, 35, 36, 38–40, 43]. However, as the currents are expected to be relatively long-lived, with electronic coherence times on scales of tens of fs that would only decay as the electron population in the valence orbitals equilibrates as a result of scattering processes, the magnetic field emission extends for several tens of fs, or longer. In order to generate shorter pulses other strategies must be employed. In what follows, we demonstrate a current-gating strategy based on non-resonant optical switches for improved temporal control over the generated ring currents, thus creating an ultrafast magnetic pulse synthesizer to precisely shape the magnetic field emission on femtosecond to attosecond scales.

We propose the optical-gating configurations depicted in Figs.2(a,b) to control the temporal profile of the ring current (Fig.2(c,d)), and thus the waveform of the magnetic field. By combining two counterrotating circularly-polarized laser pulses separated by a tunable time delay (Fig. 2(a)), we induce an ultrafast turn-on and the subsequent extinction (Fig.2(c)) of both the ring-current and the magnetic field. To completely remove the ring current, the intensity of the second pulse needs to be adjusted to a slightly higher value than the first one (blue solid line in Fig.2(c) correspond to $I_1$ =1.40 PW/cm$^2$, $I_2$ =1.55 PW/cm$^2$ at a central wavelength of 267 nm). This arises since the bound electron population depletes during the interaction, increasing the ionization potential of the system (binding the electrons more strongly), and thus, the second pulse requires a higher intensity to balance the population of $p_+$ and $p_-$ orbitals.

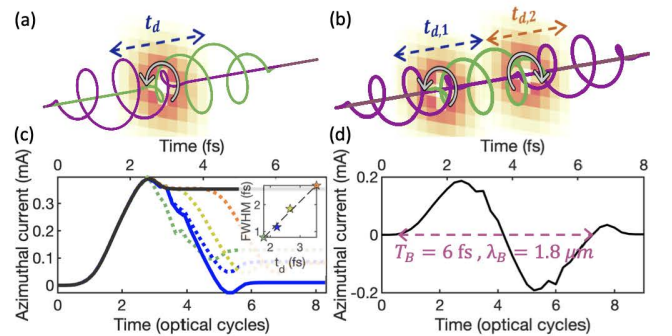

FIG. 2. Current-gating schemes for waveform control of ultrashort magnetic pulses. (a) The synthesis of two time-delayed counterrotating circularly-polarized temporally confines the emission of a D.C. magnetic pulse by limiting the temporal window of the ring current. (c) The duration of the magnetic burst can be tuned by the time delay $t_d$. The inset shows the linear relation of the full-width at half-maximum (FWHM) intensity of the magnetic field with the time delay, with the shortest pulse corresponding to 787 attoseconds. By adding a third gating pulse in (b), we synthesize in (d) the profile of a single-cycle magnetic pulse, whose wavelength can be tuned with the time delay between the pulses.

By tuning the time delay between the two pulses, $t_d$,

4we shape the waveform of the magnetic field and gain control over the pulse duration, as emphasized in the inset of Fig. 2(c), reaching sub-femtosecond duration (787 as). The full-width at half-maximum (FWHM) is defined in terms of the intensity of the magnetic pulse. Further progress in the development of high-intensity ultrashort driving pulses at shorter wavelengths could technically allow the generation of even shorter magnetic field pulses. Notably, without any additional manipulation, by setting longer time delays than the duration of the driving pulses, the magnetic field emission is temporally isolated, but they still co-propagate and overlap in space. However, optical setups can be developed to filter out the driving laser field (both electric and magnetic components) to provide full spatial and temporal separation for applications. For example, by applying spectral filtering, our controlled magnetic field emission (which is D.C. in nature and carries much lower spectral content) can be isolated from the driver and other high-frequency emissions. We should also note that circularly-polarized pulses in centrosymmetric media do not generate efficient high-harmonic emission[65–68], such that there is only a need to filter out the driving frequencies. Another option to spatially and temporally select the magnetic field emission is using polarization filtering, since the synthesized magnetic field's polarization is orthogonal to the driver. Moreover, the interaction with a noble gas avoids the additional complexity of molecular alignment, or ultrafast decoherence times due to ionic vibrations[69]. All these points should aid the experimental observation and application of our scheme.

Similar setups of collinear counterrotating circularly-polarized laser pulses are employed in polarization-gating techniques[70, 71], which are widely applied to isolate a single attosecond pulse in HHG [67, 72] when the two pulses overlap in time. Even if the physical mechanism responsible for the attosecond electric field emission in polarization gating is different than here, these experimental developments [67, 72] also benefit the implementation for generating ultrashort magnetic pulses.

To provide further tunability in the magnetic emission, we considered a train of three gating pulses as shown in Fig.2(b). Remarkably, the central wavelength of the magnetic pulse can be controlled by tuning the time delay between the laser pulses. Panel 2(d) displays the temporal waveform of a single-cycle magnetic pulse whose wavelength of 1.8 $\mu$m (corresponding to a frequency of $\sim$167 THz) is determined by a time delay of 2.2 fs. The peak intensities of the 3 pulses are set to $I_1$ =1.0 PW/cm$^2$, $I_2$ =1.4 PW/cm$^2$, and $I_3$ =1.2 PW/cm$^2$. By adjusting the peak intensities, pulse duration, and time delay of the drivers, we customize the temporal waveform of the magnetic field.

A closer inspection of Figs.2(b,d) points out that the ring current (and consequently the induced magnetic field) is maximum at the temporal gate between the two circular switch pulses (minimums of the driving electric field), which might appear counterintuitive at first glance. To clarify the physical mechanism behind this observation, we identify the role of each of the pulses in the coherent control of the ring currents. The first pulse initiates the strong-field ionization asymmetry between $p_+$ and $p_-$ valence orbitals that gives rise to a ring current whose amplitude increases during the interaction. The second pulse, due to its counterrotating circular polarization, firstly cancels this ring current, and then builds up a current flow in the opposite direction (flipping the sign of the magnetic emission) since $I_2$ is significantly higher than $I_1$. Then, the third pulse (co-rotating with the first and counterrotating with the second driver) is employed as a final optical switch to suppress the ring current. In essence, by tuning the time delay between a train of counterrotating drivers, we effectively control the time window of the magnetic emission, which oscillates following the helicity of the laser. Note also that the wavelength of the magnetic field, $\lambda_B$, is obviously determined by its period, $T_B$, by the standard relation $\lambda_B = cT_B$. In that respect, our scheme translates the time-delay parameter, which is easily tuned in experiments, to an effective spectral-domain control over the magnetic pulse. It enables the temporal and spectral shaping of the magnetic field (including its carrier wave) just by driving the system with off-resonant single-color laser fields. Hence, this approach should allow unique possibilities for exploring matter responses to variable frequency magnetic pulses. Importantly, by adding more switch pulses one could in principle synthesize any desired magnetic field temporal waveform.

Finally, we extend the study to a macroscopic magnetic source composed of several ring currents using an analytic model. This allows us to explore the expected magnetic field emission in the realistic conditions of the thin gas jets employed in experiments, whereas all results up to now only considered a single emitter. We sum the magnetic field of multiple stationary current loops placed on random positions separated by an average distance $\langle d \rangle$ (see Figs.3(a,b)), so that the longitudinal component is given by $B_z(\mathbf{r}) = \sum_\ell B_{z,\ell}(\mathbf{r})$, being $\ell$ the loop index. For the averaged static magnetic component associated with each filamentary current loop of radius $a = 0.2$ a.u. (which is roughly taken from our ab-initio TDDFT calculations), we consider the following off-axis equation[73]:

$$B_{z,\ell}(x,y,z) = \frac{B_0}{\pi\sqrt{Q}} \left( \xi(\kappa) \frac{1-\alpha^2-\beta^2}{Q-4\alpha} + \chi(\kappa) \right), \quad (3)$$

where $B_0$ is the magnetic field at the center of the loop, $\alpha = \sqrt{x_r^2 + y_r^2}/a$ and $\beta = z_r/a$ are the radial and longitudinal angles respectively defined in terms of the relative coordinates $(x_r, y_r, z_r) = (x - x_0, y - y_0, z - z_0)$, whose origin lies at the center of the loop $(x_0, y_0, z_0)$. $Q = (1+\alpha)^2 + \beta^2$ and $\kappa = \sqrt{4\alpha/Q}$ are dimensionless parameters. Then, $\chi(\kappa)$ and $\xi(\kappa)$ are the complete elliptic integral functions of the first and second kind respectively [74]. Even though this model is limited by the assumption of steady-state independent ring currents and the coherence between all the atoms in the gas, it should be a valid approximation at the peaks of the magnetic emission in our gating scheme (when the currents reach a stationary regime). Thus, our purpose is to obtain an approximation to the spatial scaling of the magnetic field under the macroscopic conditions of the experiments.

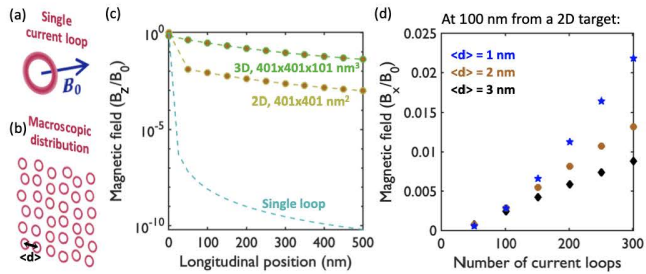

FIG. 3. Illustration of (a) a single ring-current producing a magnetic field $B_0$ at the center, and (b) a macroscopic distribution. (c) Decay of the magnetic field along the longitudinal axis for a single loop (cyan), a 2D loop distribution of 401x401 nm$^2$ (yellow), and a 3D loop target of 401x401x101 nm$^3$ (green). (d) Magnetic flux at 100 nm from the edge of the 2D distribution as a function of the number of loops for different average separation distances $\langle d \rangle$.

In Fig. 3, we analyze the decay of the macroscopic magnetic field in terms of $B_0$. The magnetic field of a single ring current drastically decreases even for small distances along the longitudinal axis (cyan line in Fig. 3(c)), which is trivially expected from the standard $1/(z^2+a^2)^{3/2}$ dependence on the Biot-Savart law. In contrast, the decay is considerably diminished for a 401x401 nm$^2$ array of current loops (yellow line in Fig. 3(c)) or even more for a volumetric distribution of 401x401x101 nm$^3$ (green line in Fig. 3(c), where the loops are separated by an average distance of $\langle d \rangle$ =2 nm). This geometry mimics realistic accessible conditions for typical strong-field ionization experiments. From a magnetostatics perspective, one could imagine an approximately infinite wall of current loops, yielding a uniform field as long as the detector is placed at a smaller longitudinal distance than the size of the wall. This provides an intuitive explanation for why the magnetic field emission is still considerable even at a reasonable distance of few hundred nanometers away from the gas jet.

This reduced spatial decay is particularly important for the applicability of such magnetic sources, since the sample can be positioned hundred of nanometers from the edge of the magnetic source, instead of requiring atomic-scale precision. Fig. 3(d) shows the magnetic flux at 100 nm from the edge of a 2D target as a function of the number of loops for different average separation distances ($\langle d \rangle$ = 1nm in blue, 2 nm in orange, and 3 nm in black). As expected, the magnetic flux increases with the ring-current density, i.e. larger number of loops and smaller separation distance. Importantly, for achivable experimental conditions, the expected magnetic flux at a distance of ∼100 nm from the gas jet can be as high as 1 T, while still generating pulses with durations of hundreds of attoseconds.

In summary, we use a train of counterrotating circularly-polarized laser pulses to control the electron population imbalance in the $p_+$ and $p_-$ valence orbitals of neon atoms. These drivers can be viewed as optical switches to temporally shape the magnetic field emission. The extension of our proposed "current-gating" technique to other systems like molecules or solids should be straightforward, since the key idea is simply to confine the magnetic emission into an ultrashort temporal window. Thus, our scheme enables the synthesis of tunable strong magnetic fields of femtosecond or attosecond pulse duration, with the possibility of isolation from the electromagnetic driver in both space and time by applying spectral or polarization filtering. Such sources constitute an ultrafast magnetic switch to probe the fastest magnetic, spin and chiral dynamics [44, 47, 48, 54–57]. Furthermore, extension to drivers in other spectral regimes (like mid-infrared or X-rays) could provide magnetic fields in the Terahertz range or at extremely high frequencies. Then, despite the extremely spatially-localized magnetic field emitted from a single ring current, the total field from multiple emitters exhibits a diminished decay of hundreds of nanometers, which can be applied in ultrafast spectroscopy experiments and attosecond metrology.

As a potential outlook, we expect that our technique could also be employed for studying the electron dynamics induced/probed by the magnetic pulse in the form of magnetic spectroscopy. By measuring magnetic fields, one could infer attosecond electron dynamics, in analogy with standard X-ray attosecond experiments or HHG spectroscopies [6, 10, 75]. We believe this would open interesting opportunities for exploring electron correlations and spin-orbit coupled dynamics in complex systems [76–80].

This work was supported by the Cluster of Excellence Advanced Imaging of Matter (AIM), Grupos Consolidados (IT1249-19), SFB925, "Light Induced Dynamics and Control of Correlated Quantum Systems". It has received funding from the European Union's Horizon 2020 research and innovation programme under the Marie Skłodowska-Curie Grant (No. 860553), the ERC Starting Grant ATTOSTRUCTURA (No. 851201), and NextGenerationEU MUR D.M. 737/2021, "Materials Manipulation with Light." The Flatiron Institute is a division of the Simons Foundation. A.H. acknowledges the financial support from Universidad de Salamanca for her international stay at MPSD. F.P.B. acknowledges financial support from the European Union's Horizon 2020 research and innovation programme under the Marie Sklodowska-Curie Grant (No. 895747, NanoLightQD). O. N. gratefully acknowledges the generous support of a Schmidt Science Fellowship.

# Supplementary Material: Tunable Tesla-scale magnetic attosecond pulses through ring-current gating

Alba de las Heras,[1,*] Franco P. Bonafé,[2] Carlos Hernández-García,[1] Angel Rubio,[2,3,4,†] and Ofer Neufeld[2,‡]

[1]*Grupo de Investigación en Aplicaciones del Láser y Fotónica, Departamento de Física Aplicada, Universidad de Salamanca, Salamanca 37008, Spain*

[2]*Max Planck Institute for the Structure and Dynamics of Matter and Center for Free-Electron Laser Science, Hamburg 22761, Germany*

[3]*Center for Computational Quantum Physics, The Flatiron Institute, New York 10010, USA*

[4]*Nano-Bio Spectroscopy Group, Departamento de Física de Materiales, Universidad del País Vasco, San Sebastián 20018, Spain*

# I. METHODS

## A. Time propagation and magnetic field emission

We perform ab-initio calculations of the laser-atom interaction in the the multi-system framework of the OCTOPUS open-access code [1, 2]. The multielectron dynamics are modeled with time-dependent density-functional theory (TDDFT) by solving in real-time and real-space the time-dependent Kohn-Sham (KS) equations[3] in the length gauge of the dipole approximation:

$$i\frac{\partial}{\partial t}|\varphi_n^{KS}(t)\rangle = \left[-\frac{1}{2}\nabla^2 + v_{KS}(\mathbf{r},t)\right]|\varphi_n^{KS}(t)\rangle \tag{1}$$

The KS potential

$$v_{KS}(\mathbf{r},t) = v_{ion} + v_H(\mathbf{r},t) + v_{xc}[n(\mathbf{r},t)] - \mathbf{r}\cdot\mathbf{E}(t) \tag{2}$$

reproduces the multi-electron density $n(\mathbf{r},t) = \sum_n |\langle\mathbf{r}|\varphi_n^{KS}(t)\rangle|^2$ associated to the KS orbitals $\varphi_n^{KS}(\mathbf{r},t)$, being $n$ the orbital index. The KS potential includes the usual terms describing the electron-nuclei interaction $v_{ion}$ (which includes also an effective core-electron shell described with norm-conserving pseudopotentials[4]), the Hartree potential $v_H(\mathbf{r},t) = \int d^3r' \frac{n(r',t)}{|\mathbf{r}-\mathbf{r}'|}$, the exchange-correlation functional $v_{xc}$ in the widely-used adiabatic local density approximation including a self-interaction correction[5], and the dipole term of the electric field interaction $-\mathbf{r}\cdot\mathbf{E}(t)$. The equations are given in atomic units (a.u.).

The electric dipole approximation is a long-established approach for modeling strong laser-matter phenomena at non-relativistic electron energies and considering laser fields whose wavelength and spatial dependence are much larger than the atomic scale. In the dipole approximation, the magnetic field of the laser pulse is neglected.

For our calculations, the equation of the total electric field $\mathbf{E}(t)$ associated with the train of counterrotating circularly-polarized laser pulses is written as a sum over $m$, the number of pulses, with electric field amplitudes $E_m$, time delay $t_d$, wavelength $\lambda$, and a 4-cycle pulse

* albadelasheras@usal.es
† angel.rubio@mpsd.mpg.de
‡ ofer.neufeld@gmail.com

duration $\tau = 8\pi/\lambda$:

$$\mathbf{E}(t) = \sum_m \frac{1}{\sqrt{2}}(\mathbf{u_x} + i(-1)^m \mathbf{u_y})E_m \sin\left(\frac{\pi(t-t_d)}{\tau}\right)^{3\left|\frac{\pi(t-t_d)}{\tau} - \frac{\pi}{2}\right|/4}$$
$$\cdot \Theta\left(\frac{\tau}{2} - |t - \frac{\tau}{2} - t_d|\right)\cos\left(\frac{2\pi c}{\lambda}t\right). \quad (3)$$

In this expression, $\Theta$ denotes the Heaviside function, and the unitary vector $\frac{1}{\sqrt{2}}(\mathbf{u_x} \pm i\mathbf{u_y})$ determines the right or left circular polarization (where $i$ is the imaginary unit). The field amplitudes $E_m$ have been varied between the range of 0.10 a.u. to 0.20 a.u, corresponding to peak intensities within 0.40 PW/cm$^2$ and 1.40 PW/cm$^2$. The driving central wavelength has been set to 800 nm, 400 nm and 267 nm, corresponding to the emission of the standard Titanium-sapphire laser (800 nm), its second harmonic (400 nm), and third harmonic (267 nm). Thus covering the infrared, visible and ultraviolet regions of the electromagnetic spectrum.

For our study, the relevant observable is the microscopic current density:

$$\mathbf{j}(\mathbf{r},t) = \frac{1}{2}\varphi_n^{KS,*}(\mathbf{r},t)\left(-i\boldsymbol{\nabla} + \frac{\mathbf{A}(t)}{c} - i[V_{ion}, r]\right)\varphi_n^{KS}(\mathbf{r},t), \quad (4)$$

where the dipole vector potential associated with the laser pulse is a time-dependent function upholding $\mathbf{A}(t) = -\int_0^\infty c\mathbf{E}(\mathbf{t})$. Concretely, in our setup of collinear circularly polarized laser pulses, the induced electronic current flows along the azimuthal direction, $\phi$, so that the important component is

$$j_\phi(\mathbf{r},t) = j_x(\mathbf{r},t)\cos(\phi) + j_y(\mathbf{r},t)\sin(\phi). \quad (5)$$

This electronic current acts as a source term in the microscopic Maxwell equations to yield a magnetic field along the light propagation direction ($z$). Therefore, synchronously to the TDDFT approach, we numerically solve the microscopic Maxwell's equations:

$$\boldsymbol{\nabla} \cdot \mathbf{E} = \frac{\rho}{\epsilon_0}; \qquad \boldsymbol{\nabla} \cdot \mathbf{B} = 0 \quad (6)$$
$$\boldsymbol{\nabla} \times \mathbf{B} = \mu_0 \mathbf{j} + \frac{1}{c^2}\frac{\partial \mathbf{E}}{\partial t}; \qquad \boldsymbol{\nabla} \times \mathbf{E} = -\frac{\partial \mathbf{B}}{\partial t}, \quad (7)$$

where $\rho$, $\epsilon_0$, $\mu_0$, and $c$ are the charge density, the vacuum electric permittivity, the magnetic permeability, and the speed of light, respectively.

Our calculations are performed in the multi-system framework of OCTOPUS[1] that uses the Riemann–Silberstein vector representation in *forward coupling*[2] (where only the incoming electromagnetic field is considered in the laser-matter interaction, thus neglecting the back-response that the electronic current could have on the electromagnetic field). This formalism is thoroughly described in section II of ref. [1], and chapters III and VI of ref. [2].

Both the KS and the Maxwell (MX) coupled solvers are numerically discretized over a cartesian grid with temporal steps of $\Delta t_{KS} = 0.01$ a.u. and $\Delta t_{MX} = 0.001$ a.u, spatial spacing $\Delta x_{KS} = \Delta y_{KS} = \Delta z_{KS} = \Delta x_{MX} = \Delta y_{MX} = \Delta z_{MX} = 0.2$ a.u., with spatial box length $L_{KS} = 50.0$ a.u., $L_{MX} = 53.2$ a.u. in each direction, using an imaginary absorbing potential of 20 a.u. width in the KS solver, and another absorbing boundary of 1.6 a.u. at the edges of the MX mesh. These parameters were chosen to guarantee the convergence and a consistent propagation.

### B. Macroscopic analytical model details

We extend our analysis to a macroscopic target composed of many ring currents in order to mimic the gas jets employed in experiments. Our analytic model for the longitudinal component of the total magnetic field, $B_z(\mathbf{r})$, considers multiple stationary independent filamentary ring currents placed on random positions and separated by an average distance $\langle d \rangle$, so that

$$B_z(\mathbf{r}) = \sum_\ell B_{z,\ell}(\mathbf{r}), \qquad (8)$$

where $\ell$ is the loop index. The radius of the current loops is set to $a = 0.2$, which is roughly the value obtained in TDDFT. Then, the off-axis magnetic field of each ring current is given by [6]:

$$B_{z,\ell}(x,y,z) = \frac{B_0}{\pi\sqrt{Q}}\left(\xi(\kappa)\frac{1-\alpha^2-\beta^2}{Q-4\alpha} + \chi(\kappa)\right), \qquad (9)$$

where $B_0$ is the magnetic field at the center of the loop, $\alpha = \sqrt{x_r^2 + y_r^2}/a$ and $\beta = z_r/a$ are the radial and longitudinal angles respectively defined in terms of the relative coordinates $(x_r, y_r, z_r) = (x - x_0, y - y_0, z - z_0)$, whose origin is randomly fixed at the center of each current loop $(x_0, y_0, z_0)$. For a more compact equation, the dimensionless parameters $Q = (1+\alpha)^2 + \beta^2$ and $\kappa = \sqrt{4\alpha/Q}$ are defined. The complete elliptic integral functions of the first and second kind[7], $\chi(\kappa)$ and $\xi(\kappa)$ respectively, are computed numerically. Overall, this model enables a rough estimation of the spatial scaling of the magnetic field emission from a macroscopic target, which in our case simulates a gas jet, but it could be applied to other systems, such as nanostructures.

## II. ADDITIONAL RESULTS

### A. Results at different gauges and without pseudopotentials

We compare the results of the time-dependent KS equations in the length and velocity gauges of the dipole approximation. Formally, both approaches should be fully equivalent, but of slightly different numerical costs. In the velocity gauge, the analogous equations are the following:

$$i\frac{\partial}{\partial t}|\varphi_n^{KS}(t)\rangle = \left[\frac{1}{2}\left(-i\boldsymbol{\nabla} + \frac{\mathbf{A}(t)}{c}\right)^2 + v_{KS}(\mathbf{r},t)\right]|\varphi_n^{KS}(t)\rangle \tag{10}$$

$$v_{KS}(\mathbf{r},t) = v_{ion} + v_H(\mathbf{r},t) + v_{xc}[n(\mathbf{r},t)] \tag{11}$$

This comparison is performed for an 8 cycles pulse, 800 nm central wavelength, 1.40 PW/cm$^2$ envelope's peak intensity, $\Delta x_{KS} = \Delta y_{KS} = \Delta z_{KS} = 0.40$ a.u., and $\Delta t_{KS} = 0.40$ a.u. The results in Fig. 1(a) obtained in the different gauges show a good agreement before the maximum of the laser pulse, but there is a small $\sim 12\%$ discrepancy in the maximum value of the azimuthal current. This discrepancy arises due to the different analytical definitions of our input external fields, based on $E(t)$ or $A(t)$ (related by a time derivative or integral). Thus, the effect is associated with a difference in the carrier-envelope phase (CEP) of the pulse. Still, we can conclude that either the length or velocity gauges are a valid approach to describe the phenomenology.

We have also investigated the influence of using pseudopotentials to describe the inner electron core-shell (1s orbital for Ne), which greatly reduces the required computational resources. To test whether this could introduce any numerical artifact, we performed an all-electron calculation test for a 4 cycles pulse, 267 nm central wavelength, 1.40 PW/cm$^2$ peak intensity, $\Delta x_{KS} = \Delta y_{KS} = \Delta z_{KS} = 0.40$ a.u., and $\Delta t_{KS} = 0.40$ a.u. In Fig. 1(b) we observe that a similar evolution is captured, apart from numerical instabilities in the all-electron calculation, likely appearing due to the employed grid spacing, which is not properly capturing the spatial structure of the 1S orbital. Such artifacts could be removed by employing tighter grid spacing. Nevertheless, we can infer from our test that the pseudopotential approach provides a proper description of the azimuthal current, and that core electron contributions are negligible.

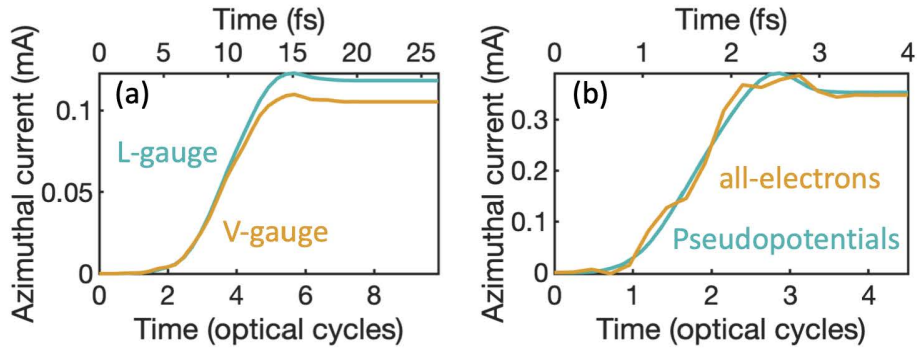

FIG. 1. Comparison of different theoretical models of the multielectron dynamics.

## B. Spatial distribution of the magnetic field from a single emitter

With our coupled KS-MX ab-initio approach, we study the interaction of a Neon atom with a driving laser pulse of 267 nm central wavelength at a peak intensity of 1.4 PW/cm$^2$, as in the main text. In the case of a single-atom emitter, the magnetic field is extremely localized. The spatial distribution of the longitudinal component, $B_z(\mathbf{r}, \mathbf{t})$, is shown in Fig. 2. The left and right panels show the dependence along the longitudinal and transversal axis respectively. The magnetic field is expressed in Teslas, whereas the spatial coordinates are given in a.u. We remark that the magnetic field shows an ultrafast turn-on and then it persists after the interaction with the laser pulse. The driving electric field is superimposed in the figures (grey line) for reference. As discussed in the main text, this magnetic field can be temporally confined and modulated by using a train of counterrotating circularly-polarized pulses. Still, the magnetic field's spatial degrees of freedom are preserved in all those cases.

Another interesting observation is that the driving laser also causes an emission of a weaker transversal circularly-polarized magnetic field component (see Fig. 3) that oscillates with the driving frequency of the laser. However, we keep in mind that these components

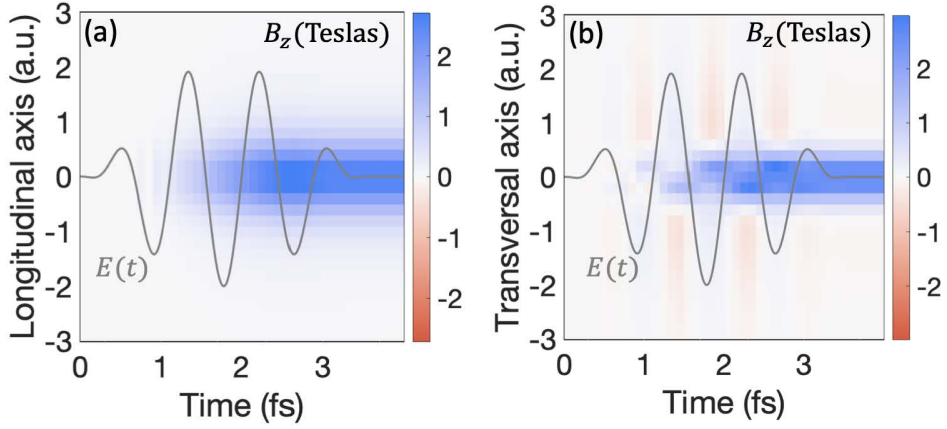

FIG. 2. Longitudinal component of the magnetic field $B_z(\mathbf{r},t)$ along (a) the longitudinal axis and (b) the transversal axis during the interaction driving laser pulse (shown in grey line in arb.u.) at 1.4 PW/cm$^2$ peak intensity and 267 nm of central wavelength. Note that the magnetic field persists after the end of the driving pulse.

spatially and temporally overlap with the driving laser, and vanish when after the laser pulse.

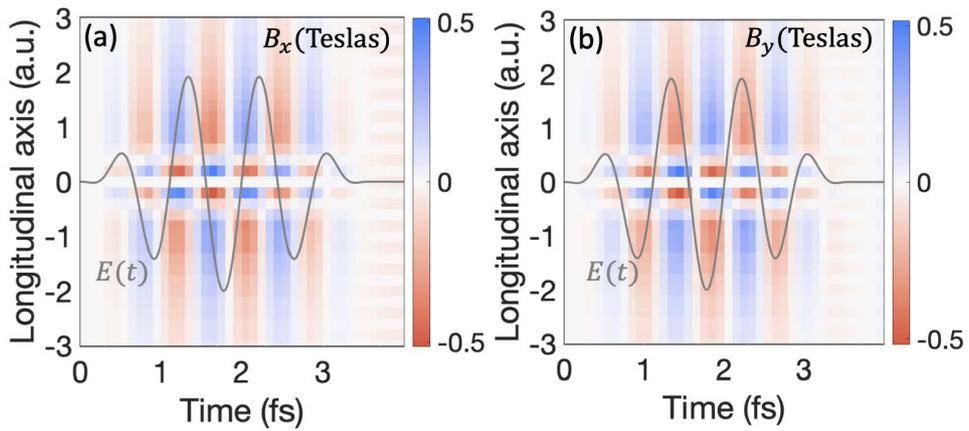

FIG. 3. Transversal magnetic field components dependence along the longitudinal axis: (a) $B_x(\mathbf{r},t)$ and (b) $B_y(\mathbf{r},t)$, during the interaction driving laser pulse (shown in grey line in arb.u.) at 1.4 PW/cm$^2$ peak intensity and 267 nm of central wavelength.